\documentclass[10pt,conference]{IEEEtran}
\usepackage[dvips]{graphicx}
\usepackage{algorithm}
\usepackage{algorithmic}

\ifCLASSINFOpdf
   \usepackage[pdftex]{graphicx}
\else
   \usepackage[dvips]{graphicx}
   \graphicspath{{./eps/}{./jpeg}}
   \DeclareGraphicsExtensions{.eps,.pdf}
\fi
\begin{document}
\title{ E-DTN : A Multi-Interface Energy DTN Gateway}
\author{\IEEEauthorblockN{Prabhakar T.V, Akshay Uttama Nambi S.N, Jamadagni H.S}
\IEEEauthorblockA{Centre for Electronics Design and Technology,
Indian Institute of Science, Bangalore, India\\
(tvprabs,akshay,hsjam)@cedt.iisc.ernet.in}
}
\IEEEoverridecommandlockouts
\IEEEpubid{\makebox[\columnwidth]{978-1-4244-8953-4/11/\$26.00~\copyright~2011 IEEE \hfill} \hspace{\columnsep}\makebox[\columnwidth]{ }}
\maketitle
\begin{abstract}
To overcome the problem of unavailability of grid power in rural India, we explore the possibility of powering WSN gateways using a bicycle dynamo.  The ``Data mule'' bicycle generates its own power to ensure a self sustainable data transfer for information dissemination to small and marginal farmers. Our multi-interface WSN gateway is equipped with Bluetooth, Wi-Fi and GPRS technologies.  To achieve our goal, we exploit the DTN stack in the \emph{energy sense} and introduce necessary modifications to its configuration.\\
\end{abstract}
%
\IEEEpeerreviewmaketitle
\section{Introduction}
\begin{figure}
\centering
\includegraphics[width=3.0in,height=2.2in]{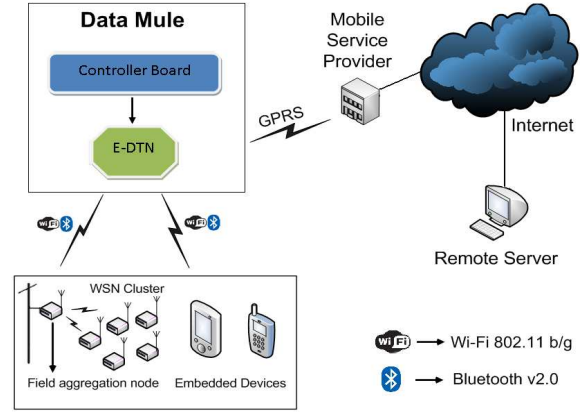}
\caption{ Block diagram of E-DTN multi-interface.}
\label{fig:bigpicture}
\end{figure}

Several gateway technologies exist today to relay data aggregated from an ad-hoc sensor network cluster. Such technologies include Bluetooth, Wi-Fi, and GSM/GPRS. While GPRS has the added advantage of relaying the data directly to the Internet, Bluetooth and Wi-Fi can be used to relay data over short to medium range respectively. One deterrent to the wide-spread use of such technologies in the rural context comes from the fact that most villages in India have very little access to grid power. Often power cuts last for 12-16 hours a day. GPRS technology requires sufficiently high energy with peak currents of about $1.6$A during data transmissions. Even large battery backups are insufficient to guarantee its continuous operation. Is there a solution to this problem? Can we generate power just sufficient for GPRS transmission? Our work positions itself to tackle the issue of powering the GPRS gateway from harvested energies. Fig.\ref{fig:bigpicture} shows the block diagram of E-DTN multi-interface gateway.  Alongside are shown sensor network gathering data in the field and other embedded devices such as camera phones and data modems. In the agriculture context, the purpose of a sensor network deployment is to collect data to provide information to small and marginal farmers about the standing crop by evaluating its stress in adverse situations such as drought and pest attacks that impact the yield.  The requirement for data gateway is to relay data for the purpose of analysis and decision science.
\section{Goal and Related work}
Our goal is to demonstrate the capabilities of a grid independent hybrid data relay communication system comprising of Bluetooth, and Wi-Fi.  GPRS technology is used as the Internet gateway.  We use a bicycle dynamo to generate this energy. In this energy generating system, data downloads are possible over Wi-Fi or Bluetooth,  and upload to Internet uses GPRS technologies. Since energy is generated on the fly, it now becomes necessary to negotiate this quantity. In our work, we employ the Delay/Disruption Tolerant Network (DTN) stack and exploit its features from the view of \emph{energy availability} rather than connectivity and we therefore call our system ``E-DTN''. In \cite{throwbox} the authors discuss an energy driven system to improve packet delivery in a sparse sensor deployment. In this work, we adapt packet buffering and propose an algorithm towards an energy based data transfer, where data bundles are exchanged between E-DTN end points to match the minimum energy available between the node pairs.  Thus our scheme is comprehensive compared to \cite{throwbox}. Using E-DTN, energy availability in terms of ``energy bundles'' is negotiated.  The input parameters considered for negotiation include: (a) energy availability (b) data rate (c) transmit power and finally (d) channel state based on signal strength. The outcome determines the data transfer either over Bluetooth or Wi-Fi. We show that energy stored in a super capacitor is sufficient for our purpose.
\begin{figure}
\centering
\includegraphics[width=3.0in,height=1.9in]{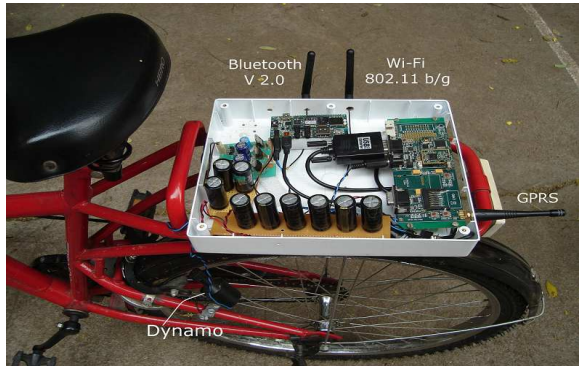}
\caption{Dynamo Driven Data mule.}
\label{fig:bicycle_dtn}
\end{figure}
\section{Implementation}
\begin{figure}
\centering
\includegraphics[width=3.0in,height=1.8in]{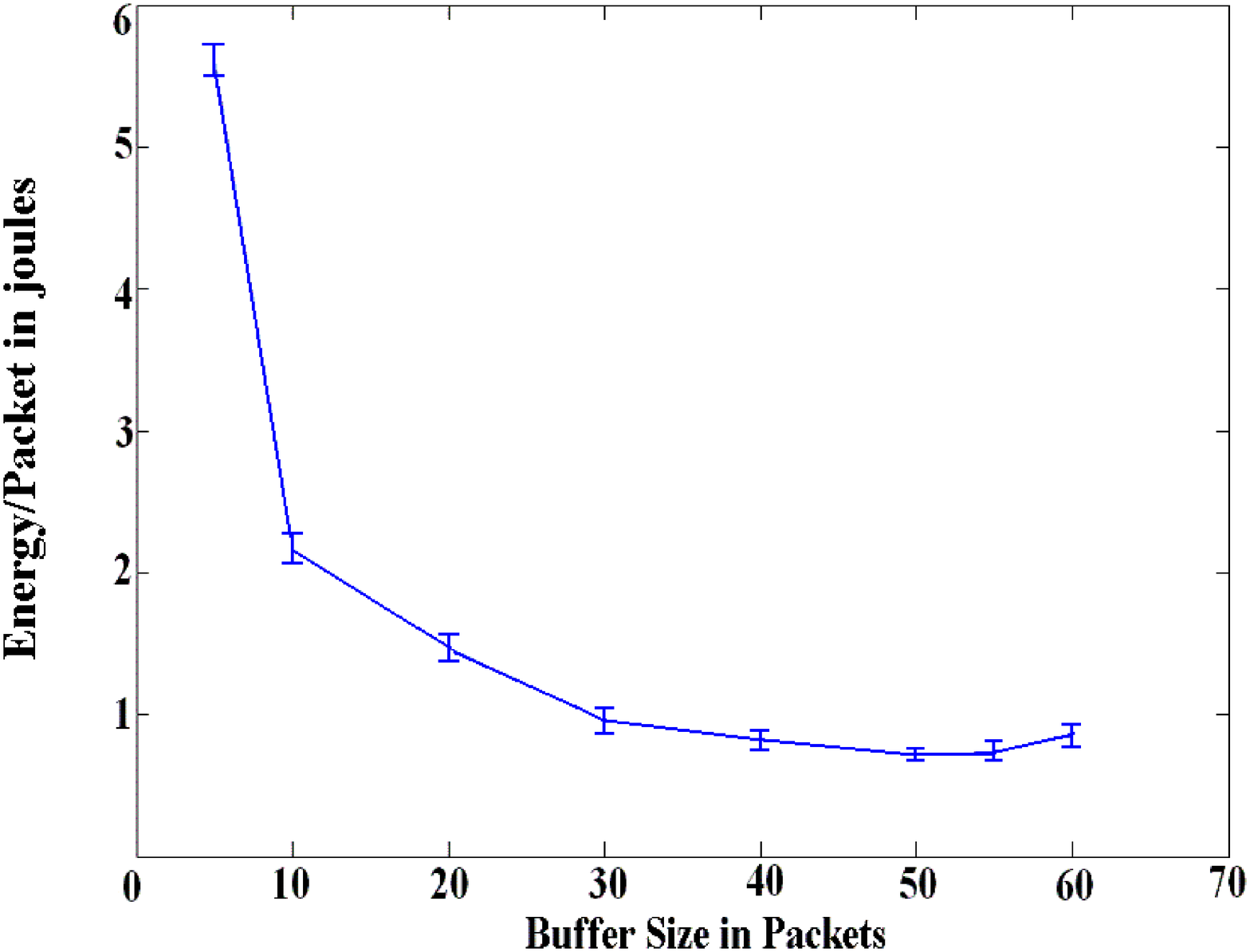}
\caption{Energy per packet vs Buffer size for 95\% Confidence Interval}
\label{fig:confidence}
\end{figure}

The initial latency for the energy bundle transfer is $6$ seconds from the time the Data mule (DM) and Field Aggregation Node (FAN) estimate their energies. We implemented the Data mule using Gumstix's system on module (SOM) Overo Fire as the controller, and Siemens TC65 as the GPRS module. 
Table \ref{energywifi} shows the split time and energy break-up for a single bundle transfer. The data mule consumes around $190$s and $360$J for a bundle transfer using DTN over Wi-Fi to download the data from FAN and GPRS for uploading the bundle to the server. Fig.\ref{fig:bicycle_dtn} shows these  system components including the super capacitor banks for storing the energy.\\ 
Table \ref{energy} shows the latency in a single bundle transfer between the E-DTN end points over Bluetooth and Wi-Fi. We experimentally evaluated the optimal size of the GPRS buffer. The packet size was fixed at $32$ bytes. Experiments were conducted by varying the buffer size and programming the GPRS module. Once the buffer is full, the GPRS radio is switched to ``on'' state. 
Fig. \ref{fig:confidence} shows 95\% confidence interval of Energy/Packet to transmit. As we increase the buffer size on the module, the transfer energy for a packet decreases until the buffer size is $50$\,packets. Soon, the energy increases, although very slowly. By taking the $50$\,packet buffer, our results show that in order to complete a GPRS transfer for a single packet, the minimum amount of energy consumed is $700$\,mJ. The packet delivery latency is $31$s and energy consumed is $35$\,J  for $50$\,packets.

\begin{table}
\centering
\caption {single bundle transfer over Wi-Fi}
\label{energywifi}
\begin{tabular}{|l|c|c|} \hline
Operations & Energy [J] & Time [s] \\\hline
Powering up the SOM and GPRS module&77&30\\\hline
Auto-login on SOM&86&30\\\hline
DTN communication( Send and receive )&42&13\\\hline
Bundle transfer from SOM to GPRS &42&14\\\hline
SOM shutdown &60&15\\\hline
Siemens TC65 startup &25&60\\\hline
GPRS transmission to the server&35&31\\\hline
\end{tabular}
\end{table}
\begin{table}
\centering
\caption {Latency for one DTN Bundle transfer over Bluetooth and Wi-Fi}
\label{energy}
\begin{tabular}{|l|c|c|} \hline
Bundle size & Time taken over & Time taken over\\
  &  Bluetooth in s &  Wi-Fi in s \\ \hline
5 kB&5&7\\\hline
1 MB&90&7\\\hline
3 MB&280&20\\\hline
\end{tabular}
\end{table}
\section{Results and Conclusion}
We used a $100$\,F super capacitor across all our measurements. Since our energy requirement is to the extent of retrieving the data bundles from the FAN and transferring the same over a GPRS link, we do not require an infinite buffer. Based on the energy measurements we conducted, A $75$F capacitor is sufficient to transfer one data bundle of $50$ packets over GPRS. The advantage of this optimal value ensures that the cyclist does not have to pedal for longer periods to kick-off packet transmissions. We found that $22$\,minutes of cycling at about $13$\,kmph is required to generate energy sufficient to transfer a $50$ packet buffer. We measured $2.9$ watts as the power generated from the dynamo.\\
The model we have proposed becomes sustainable and general enough for application in several scenarios. It is sustainable in the field due to the fact that there are no replaceable components such as batteries and associated charging electronics. An ideal super capacitor has infinite charge-discharge cycles and does not require complex charging circuitry. Thus, DTN from an energy perspective combined with reliability is a novelty in our proposed scheme. The solution is general enough for application in future home networks as well, where home networks require zero downtime. The only way to ensure this in today's world is to make users generate their own power.
\begin{algorithm}
\caption{E-DTN bundle transfer algorithm}\label{algo1}
\begin{algorithmic}[1]
\STATE
The Field Aggregation Node (FAN)  and the Data Mule (DM) estimate their respective energies [$E_F$] and [$E_D$].
\STATE
A decision is made to determine the number of bundles ``n'' to be sent from the FAN based on the minimum energy i.e., $n$ = min($E_D$,$E_F$).
\STATE
The outcome of the energy negotiation determines the number of bundles and the technology (Wi-Fi or Bluetooth) to be used for the data transfer.
\STATE
The FAN transmits the bundles to the DM and awaits an ACK.
\STATE
The DM, after receiving a bundle, starts the GPRS buffer transmission. On successful delivery at the remote server, the DM sends an ACK back to the FAN. 
\STATE
The FAN deletes all successfully acknowledged bundles.
\STATE
If the DM has more energy, it puts up a fresh request and Steps from 2 to 7 are repeated.
\end{algorithmic}
\end{algorithm}
\bibliographystyle{IEEEtran}
%

\end{document}